\documentclass{ws-procs9x6-cpt16}

\begin{document}

\newcommand{\refeq}[1]{(\ref{#1})}
\def\etal {{\it et al.}}

\title{Towards Sympathetic Laser Cooling and\\ Detection of Single (Anti-)Protons }

\author{T.\ Meiners,$^1$ M.\ Niemann,$^1$ A.-G.\ Paschke,$^{1,2}$ 
M.\ Borchert,$^1$ A.\ Idel,$^1$ J.\ Mielke,$^1$ \\
K.\ Voges,$^1$ A.\ Bautista-Salvador,$^{2,1}$ R.\ Lehnert,$^{3,1}$ 
S.\ Ulmer,$^4$ and C.\ Ospelkaus$^{1,2}$}

\address{$^1$Institut f\"ur Quantenoptik, Leibniz Universit\"at Hannover\\
	Welfengarten 1, 30167 Hannover, Germany}
\address{$^2$Physikalisch-Technische Bundesanstalt\\
	Bundesallee 100, 38116 Braunschweig, Germany}
\address{$^3$Indiana University Center for Spacetime Symmetries\\
	Bloomington, IN 47405, USA}
\address{$^4$Ulmer Initiative Research Unit, RIKEN\\
	Hirosawa, Wako, Saitama 351-0198, Japan}

\begin{abstract}
Current experimental efforts to test the fundamental CPT symmetry with single (anti-)protons are progressing at a rapid pace but are hurt by the nonzero temperature of particles and the difficulty of spin state detection. We describe a laser-based and quantum logic inspired approach to single (anti-)proton cooling and state detection. 
\end{abstract}

\bodymatter

\section{Introduction}

Penning-trap based precision measurements have been able to place stringent bounds on CPT violation such as by comparing the magnetic moment or $g$-factor of the electron and the positron,\cite{dehmelt_experiments_1990} and further improvements are anticipated.\cite{hanneke_new_2008} In these experiments, the particle is detected via the image charge induced in the trap electrode by the motion of a single particle. The temperature of the particles is related to the temperature of the cryogenic tank circuit employed in the image charge detection. The spin degree of freedom can be measured using the continuous Stern-Gerlach effect.\cite{dehmelt_proposed_1973} For heavier particles such as the proton and antiproton, trap frequencies tend to be much lower, and cryogenic cooling typically cannot be used to reach the motional ground state. Furthermore, the continuous Stern-Gerlach effect is proportional to $\mu/m$, where $\mu$ is the magnetic moment of the particle and $m$ its mass, making it much more difficult to detect the spin states.\cite{mooser_direct_2014} Heinzen and Wineland\cite{heinzen_quantum-limited_1990} proposed a method set to detect, cool and manipulate a charged particle of interest through a laser-cooled ion. The coupling is provided through the image charges induced by the motion of both particles in a common trap electrode. Another method is to couple the two particles in a double-well potential.\cite{wineland_experimental_1998} Here, we focus on the latter approach. 

\begin{figure}
	\centering
	\includegraphics[width=1.0\columnwidth]{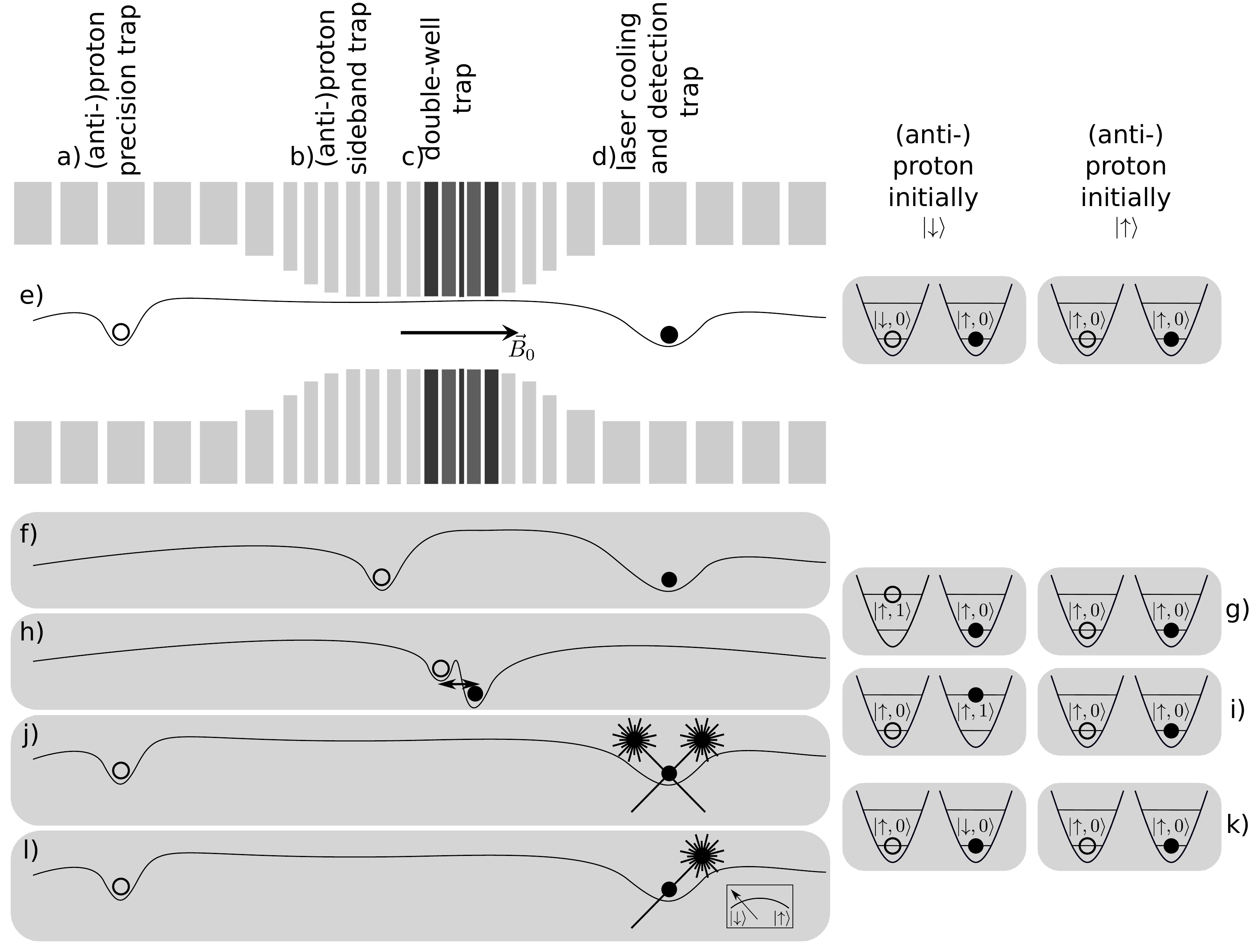}
	\caption{Conceptual Penning trap array (top) and procedure used for quantum logic inspired spin state detection of single (anti-)protons.}
	\label{aba:fig1}
\end{figure}

\section{Measurement protocol}
Experiments will be conducted in a cryogenic Penning trap array adapted from the BASE experiment.\cite{smorra_base_2015} Fig.\ \ref{aba:fig1} (top) shows a cut along the symmetry axis of a multi-zone cryogenic Penning-Malmberg trap with static magnetic field $B_0=5\,\mathrm{T}$. The trap consists of at least four individual zones (a-d), each with a dedicated function explained below. We will start to explain how the spin state of the (anti-)proton can be measured using quantum logic inspired operations, and then expand to describe a full $g$-factor measurement. Throughout the discussion, we will talk about the proton, with the understanding that, unless explicitly mentioned, the identical method can be applied to the antiproton. 

We will assume that the proton's (open circle) axial motion has been sympathetically cooled to the ground state using a laser-cooled $^9$Be$^+$ ion\cite{brown_coupled_2011} exploiting the double-well trap of Fig.\ \ref{aba:fig1}c. We start with the proton (open circle) in the precision trap (Fig.\ \ref{aba:fig1}a) and the $^9$Be$^+$ ion (filled circle) in the laser cooling and detection trap (Fig.\ \ref{aba:fig1}d). We will assume that the spin state of the proton is unknown, but $^9$Be$^+$ has been initialized in $\left|\uparrow\right>=\left|S_{1/2},m_I=3/2,m_J=1/2\right>$ through optical pumping. The two columns on the right-hand side illustrate the quantum states of the two particles throughout the procedure for the proton initially in $\left|\downarrow\right>$ (left) or in $\left|\uparrow\right>$ (right). In order to measure the proton's spin state, it will first be shuttled into the so-called proton sideband trap (Fig.\ \ref{aba:fig1}b) by applying voltage ramps to the trap electrodes. An rf blue motional sideband pulse (Fig.\ \ref{aba:fig1}f) will map the proton spin states $\left|\downarrow\right>$, $\left|\uparrow\right>$ into the motional states $n=1$ and $n=0$ (Fig.\ \ref{aba:fig1}g). This pulse can be realized using different techniques\cite{mintert_ion-trap_2001,ospelkaus_trapped-ion_2008} already demonstrated experimentally with atomic ions\cite{ospelkaus_microwave_2011,khromova_designer_2012} in the context of quantum logic. Next, the proton and the $^9$Be$^+$ ion will be shuttled to separate, but near-by potential wells of the double-well trap (Fig.\ \ref{aba:fig1}h). In this specially tailored potential with equal trap frequencies for both particles, the two charges interact remotely via the Coulomb interaction. Over one motional exchange period, the motional state of the two particles will have swapped (Fig.\ \ref{aba:fig1}i). Note that we show a potential suitable for two positively charged particles; for the antiproton, one of the dips will be inverted. In the context of quantum information processing with trapped ions, this double-well technique has been shown with pairs of atomic ions.\cite{brown_coupled_2011,harlander_trapped-ion_2011} Subsequently, the $^9$Be$^+$ ion will be shuttled back into the laser cooling and detection trap. Using a laser-induced stimulated Raman blue sideband transition (Fig.\ \ref{aba:fig1}j), the conditional motional excitation can be mapped back into the spin state of the $^9$Be$^+$ ion (Fig.\ \ref{aba:fig1}k). The net result is that the initial spin state of the proton has been fully transferred to the $^9$Be$^+$ ion (Fig.\ \ref{aba:fig1}k) and can be measured using laser-induced resonance fluorescence. By shining in a resonant laser beam, connecting the $^9$Be$^+$ $S_{1/2}$ and $P_{3/2}$ levels, the ion can be made to scatter photons if and only if it is in $\left|\uparrow\right>$. Therefore, the $^9$Be$^+$ ion will appear as a bright spot on the detector in case the proton was initially in $\left|\downarrow\right>$, and dark otherwise. One can thus determine the previous spin state of the proton using quantum logic operations and initialize the proton in $\left|\uparrow\right>$, independent of its initial state. 

To make this a full Larmor frequency measurement, one applies an rf drive at a frequency $f$ near the expected proton spin flip (Larmor) frequency $f_\mathrm{L}$ in the proton precision trap and repeats the entire detection process. By varying the drive frequency $f$ and repeating the full sequence, the transition probability can be measured as a function of $f$ and the Larmor frequency $f_\mathrm{L}$ can be determined. Together with a similar procedure to measure the proton motional frequencies, the free cyclotron frequency $f_\mathrm{c}$ can be extracted,\cite{brown_geonium_1986} and the $g$-factor is given by $g=2f_\mathrm{L}/f_\mathrm{c}$.

\section*{Acknowledgments}
We acknowledge discussions with members of the BASE collaboration 
and the NIST ion storage group. 
We acknowledge financial support from ERC StG ``QLEDS,'' 
DFG through CRC 1227 (DQ-mat), project B06, QUEST,
the Alexander von Humboldt Foundation, 
and Leibniz Universit\"at Hannover. 
We are grateful for support by the PTB clean room facility team.

\end{document}